\documentclass[prl,amsmath,amssymb,twocolumn,superscriptaddress,showpacs]{revtex4}

\usepackage{amsmath,amssymb}
\usepackage[usenames]{color}
\usepackage{amssymb}
\usepackage{grffile}
\usepackage[pdftex]{graphicx}
\usepackage{amsmath, amstext, amssymb, amsfonts, amsxtra}
\usepackage{textcomp}
\usepackage{xspace}
\usepackage{color}

\newcommand\e{{\rm e}}
\newcommand\ii{{\rm i}}
\newcommand\HHop{\mathcal{H}}

\newcommand{\hannover}{Institut f\"ur Theoretische Physik, Leibniz Universit\"at Hannover, Appelstr. 2, DE-30167 Hannover, Germany}

\begin{document}

\title{Engineering interactions and anyon statistics by multicolor lattice-depth modulations}

\author{Lorenzo Cardarelli} 
\affiliation{\hannover}
\author{Sebastian Greschner} 
\affiliation{\hannover}
\author{Luis Santos}
\affiliation{\hannover}  

\date{\today}

\begin{abstract}
We show that a multicolor modulation of the depth of an optical lattice allows for a flexible independent control of correlated hopping, occupation-dependent gauge fields, effective on-site 
interactions without Feshbach resonances, and nearest-neighbor interactions. As a result, the lattice-depth modulation opens the possibility of 
engineering with minimal experimental complexity a broad class of lattice models in current experiments with ultra-cold atoms, including Hubbard models with correlated hopping, peculiar extended models, and 
two-component anyon-Hubbard models.
\end{abstract}



\maketitle


Floquet engineering -- the averaging of fast periodic modulations to obtain an effective time-independent system --  is an ubiquitous tool for the manipulation and probing of 
various systems, ranging from NMR probes in solid state physics to atom-light interactions or Raman-dressed states~\cite{Haenggi98}. 
In recent years, Floquet techniques have established themselves as a toolbox for the creation of novel Hamiltonians for ultra-cold atoms in optical lattices, 
including lattice shaking~\cite{Eckardt05,Lignier07,Kierig2008,Zenesini2009,Struck2011,Struck2012,Parker2013}, Raman-assisted 
hopping~\cite{Aidelsburger2013,Miyake2013,Mancini2015,Stuhl2015}, and modulated interactions~\cite{Gong2009,Abdullaev2010,Rapp2012,DiLiberto2014,Greschner14doubleshaking,Meinert2016}.

A major reason for the interest on Floquet techniques lies in the possibility of engineering gauge fields, i.e. complex hopping rates, for neutral atoms in optical lattices~\cite{Struck2012}. 
Most relevantly, synthetic magnetic fields have been created in the last years using Raman-assisted hopping~\cite{Aidelsburger2013,Miyake2013,Mancini2015,Stuhl2015}. 
Interestingly, various Floquet techniques have been recently proposed for the creation of occupation-dependent gauge fields~(ODG)~\cite{Greschner14AB,Keilmann2011,Greschner15AHM,Straeter2016}, 
in which the phase of the hopping depends on the site occupation. Under proper conditions, 1D models with ODG may be mapped into an anyon-Hubbard model~(AHM)~\cite{Greschner14AB,Keilmann2011,Greschner15AHM,Straeter2016}, in which the exchange statistics of the atoms may be externally modified. The 1D AHM 
presents a wealth of new physics, including statistically-induced transitions~\cite{Keilmann2011}, novel superfluid 
phases~\cite{Greschner15AHM}, smooth fermionization~\cite{Straeter2016}, asymmetric momentum distributions~\cite{Hao2009,Tang2015}, 
and intriguing dynamics~\cite{DelCampo2008,Hao2012,Wang2014}. 
The atomic back-action on the synthetic gauge field given by ODG  
could pave a way for the realization of dynamical gauge fields~\cite{Wiese2013, Bermudez2015}, and leads to interesting physics, such as chiral solitons in 
Bose-Einstein condensates~\cite{Edmonds2013} or density-flux interplay in 2D lattices~\cite{Greschner15DDSM}.

In this Letter we propose a novel method based on the multi-color modulation of the depth of a tilted optical lattice. As shown by Ma et al.~\cite{Ma2011} 
lattice-depth modulations may be employed to assist different occupation-dependent hoppings for sufficiently strong interactions. 
We show for the particular case of two-component fermions that a three-color modulation~(3CM) of the lattice depth may be employed to achieve a separate flexible control 
of correlated hopping, ODG, effective on-site interactions without the need of Feshbach resonances, and nearest-neighbor~(NN) interactions. As a result, 3CM allows 
with a minimal experimental complexity for engineering a broad class of lattice Hamiltonians using ultra cold atoms, including Hubbard models with correlated hopping, 
peculiar extended models, and two-component AHM, whose basic properties we analyze as well.



\begin{figure}[t]
\begin{center}
\includegraphics[width=1\linewidth]{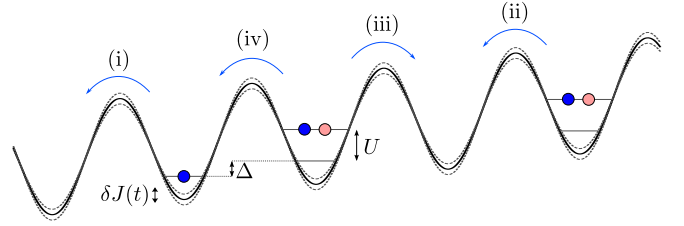}
\caption{(Color online) Sketch of the lattice set-up and the relevant hoppings.}
\vspace*{-0.5cm}
\label{fig:scheme}
\end{center}
\end{figure}




\paragraph{Effective Hamiltonian.--} We consider a balanced two-component ($\sigma=$ $\uparrow$,  $\downarrow$) Fermi gas in an optical lattice (equal for both components), whose depth is modulated in time, $V(t)=V_0+\delta V(t)$, 
with $\delta V\ll V_0$. We choose two-component fermions for simplicity, but similar ideas may be applied 
to bosons, and multi-component fermions. In the tight-binding regime, the hopping rate is $\frac{J(s)}{E_{rec}}=\frac{4}{\sqrt{\pi}} s^{3/4} \exp (-2\sqrt{s})$~\cite{Zwerger2003}, 
where $s=V/E_{rec}=s_0+\delta s(t)$, with 
$E_{rec}$ the recoil energy associated to the laser that creates the lattice. Since $\delta s\ll s_0$, then $J(t)=J_0+\delta J(t)$, where $J_0=J(s_0)$, and 
$\frac{\delta J(t)}{J_0}=\left ( \frac{3}{4}-\sqrt{s_0} \right ) \frac{\delta s(t)}{s_0}$, and hence the lattice modulation directly maps into a modulation of the 
hopping rate. We assume a tilted lattice, with an energy shift $\Delta$ between neighboring sites~(Fig.~\ref{fig:scheme}). 
The system is then described by the Fermi-Hubbard Hamiltonian:
\begin{equation}
\HHop(t)\!=\! -J(t)\sum_{j,\sigma} \! \left [ c_{j+1,\sigma}^\dag c_{j,\sigma}\! +\! \mathrm{H.c.} \right ]\! + U \HHop_{\rm int} + \Delta \HHop_{\rm tilt},
\label{eq:H_full_J(t)}
\end{equation}
where $c_{j,\sigma}$ is the annihilation operator of a fermion with spin  $\sigma$ at site $j$, 
$U$ characterizes the on-site interactions, $\HHop_{\rm int} \!=\! \sum_j n_{j,\uparrow}n_{j,\downarrow}$, and 
$\HHop_{\rm tilt} \!=\! \sum_{j,\sigma} j n_{j,\sigma}$. Note that four different hoppings are possible~(Fig.~\ref{fig:scheme}): 
(i) a single atom hops to an empty site to its right leading to an energy shift $\Delta E_1\!=\!\Delta$; 
(ii) an atom with spin $\sigma$, initially alone at a given site, tunnels to the site at its right already occupied by 
a single atom with $\bar\sigma\neq\sigma$, resulting in a shift $\Delta E_{2}\!=\!\Delta\!+\!U$; 
(iii) same as (ii) but the hopping is to the left -- in this case $\Delta E_{3}\!=\! U\! -\!\Delta$; 
and (iv) an atom of component $\sigma$ sharing a site with a $\bar\sigma$ atom, tunnels into the site at its right already occupied by 
a single atom with $\bar\sigma$ leading to $\Delta E_{4}\!=\!\Delta$ (i.e. (iv) and (i) are resonant).

We assume that $J(t)\ll \Delta, |\Delta\pm U| $, and hence direct hopping is negligible. However, a periodic modulation of $\delta J(t)$
leads to assisted hopping if the modulation frequency matches the energy shift associated to the hopping process~\cite{Ma2011}. 
Crucially, processes (i), (ii) and (iii) are characterized by different energy shifts~(typically separated by several kHz, see below), 
and hence the different hoppings may be laser-assisted separately. The key point of our proposal is to address them separately but simultaneously using a 3CM 
of the laser intensity: $\delta V(t)=\sum_{s=1,2,3} \delta V_s \cos(\omega_s t+\phi_s)$, which, as mentioned above, translates into an equivalent modulation of the hopping, 
$\delta J(t)=\sum_s \delta J_s \cos(\omega_s t+\phi_s)$. Each component of the modulation has an amplitude $\delta J_s$ and a dephase $\delta\phi_s$, which may 
be independently controlled.  The frequencies $\omega_1=\Delta$, $\omega_2=\Delta+U-\tilde U$, and $\omega_3=-\Delta+U-\tilde U$, with 
$|\tilde U|\ll U$ are chosen (quasi-)resonant to the hoppings (i) (and hence also (iv)), (ii), and (iii), respectively.

In interaction picture, $\tilde\HHop=\mathcal{U}^\dag \HHop \mathcal{U}$, with $\mathcal{U}=\exp\left [-\ii t( \Delta \HHop_{\rm tilt} + U \HHop_{\rm int} )\right ]$:
\begin{equation}
\!\!\tilde\HHop(t) \! = \! J(t) \! \sum_{j,\sigma}\! \left [ c_{j,\sigma}^\dagger 
\e^{{\rm i}t \left[\Delta + U (n_{j,\bar{\sigma}} -n_{j+1,\bar{\sigma}})\right]}  
c_{j+1,\sigma}\! +\! {\rm H.c.} \!\right ]\!.
\label{eq:H_trans_inv}
\end{equation}
3CM introduces oscillating terms $e^{\pm \ii(\omega_s \pm \Delta E_{s'})t}$. For $|\Delta-U|,U \gg J_0$ 
the fast-oscillating terms average to zero~(rotating wave approximation~(RWA)), and only quasi-resonant terms remain~\cite{Suppl}. As a result, 
processes (i)~(and (iv)), (ii), and (iii) present an effective hopping rate $\frac{\delta J_s}{2} e^{i\phi_s}$, with $s=1$, $2$, and $3$, respectively. 
We consider below the particular case with $\delta J_{2,3}=\beta\,\delta J_1$, $\phi_1=0$, $\phi_{2,3}=\phi$. 
Undoing the interaction picture we obtain the effective time-independent Hamiltonian:
\begin{equation}
\!\HHop_{eff}\!=\! -\frac{\delta J_1}{2} \!\sum_{\sigma,j}\! c_{j+1,\sigma}^\dag F[ |n_{\bar\sigma, j+1}\!-\! n_{\bar\sigma,j}|] c_{j,\sigma} + \tilde U \HHop_{int}, 
\label{eq:H_eff}
\end{equation}
where $F[0]=1$, and $F[1]=\beta e^{i\phi}$. 
3CM provides remarkable control possibilities. Both the amplitude and the phase of the hopping rate of the $\sigma$ component 
depend on the site occupation of the $\bar\sigma$ component. As shown below, this may be employed to realize ODG. 
Moreover, the detuning $\tilde U$ results in an effective on-site interaction, allowing for controlling interactions 
even in those systems where Feshbach resonances are not available. This is in particular the case of alkaline-earth fermions in the lowest $^1$S$_0$ state~\cite{Cazalilla2014}. 
Since 3CM may be also used with multi-component fermions, this opens a novel way of controlling the properties of SU(N) fermions~\cite{Cazalilla2014}.



\begin{figure}[t]
\begin{center}
\includegraphics[width=1\linewidth]{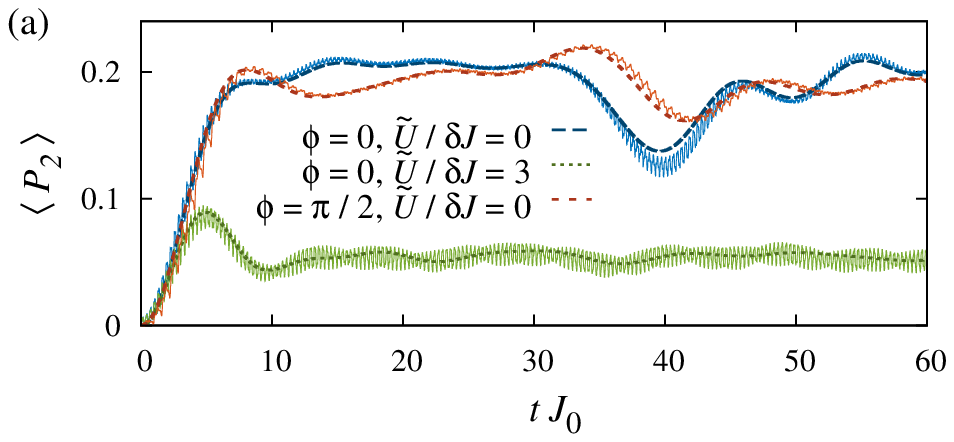}
\includegraphics[width=1\linewidth]{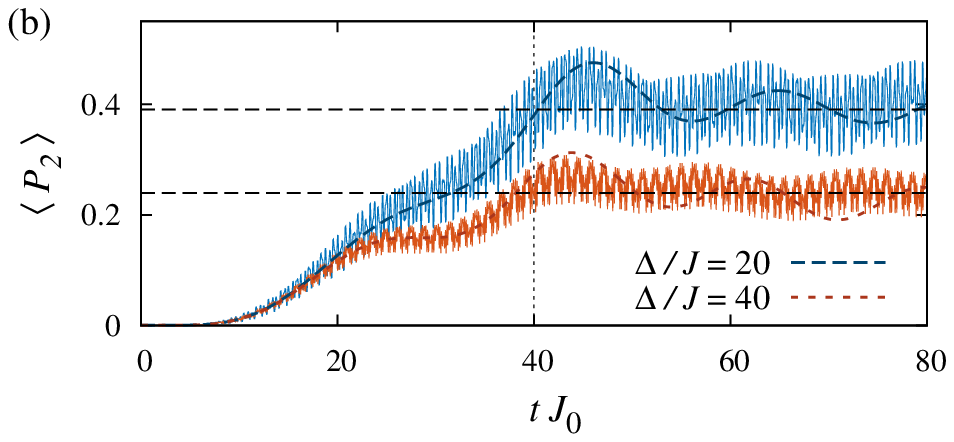}
\caption{(Color online) (a) Average double occupancy $\langle P_2 \rangle$ after a sudden-quench of $\delta V$ for a finite temperature $T=J_0$, 
$\Delta/J_0=20$, $U/J_0 = 10$, $\delta J_1/J_0 = 0.2$, $\beta=1$, and different values of $\tilde{U} / \delta J_1$ and $\phi$~(exact diagonalization results using $6$ particles in $6$ sites); 
dashed~(solid) curves depict the results of the effective~(full) model;  
(b) $\langle P_2 \rangle$ for a quasi-adiabatic preparation~(iTEBD results for $\rho=1$). The system is initially prepared in a MI for $\delta V=0$.  
$\delta V(t)$ is linearly increased to its final value for $0<J_0 t<40$ ; we consider $U /J_0 = 5$, $\delta J_1 / J_0= 0.1, \tilde{U}/\delta J_1=2$, $\beta=1$. 
$\langle P_2\rangle(t)$ for the full~(solid) and effective model~(dashed) oscillates around the expected value~(horizontal lines) for the ground state with the final $\delta V$.}
\label{fig:2}
\end{center}
\end{figure}



\begin{figure*} [t]
\centering
\includegraphics[width=1\linewidth]{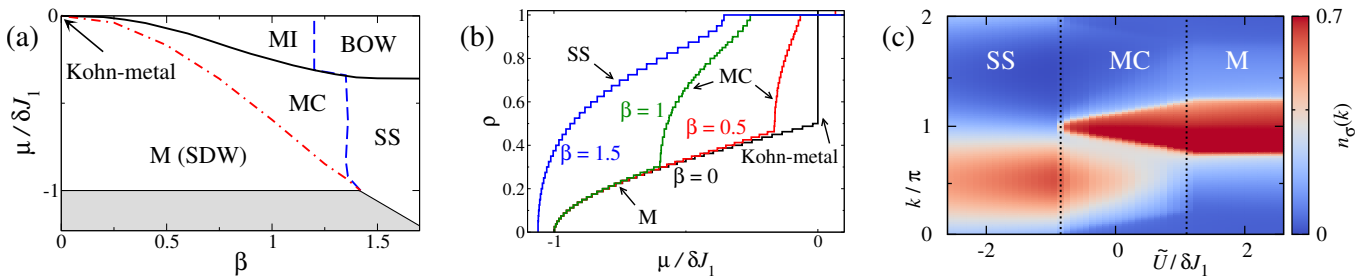}
\caption{(Color online) (a) Phase diagram of $\HHop_{eff}$ as function of $\mu/\delta J_1$ and $\beta$ for $\phi = \pi/2$ and $\tilde{U}=0$~\cite{footnote-density}. 
The dash-dotted lines mark the C-IC M-MC transition. 
The dashed~(blue) lines denote the opening of $\Delta_S$ that marks the MC-SS and MI-BOW transitions. 
Shaded regions depict the vacuum.
(b) Equation of state $\rho = \rho(\mu)$ for $\beta=0$, $0.5$, $1$, and $1.5$ for the parameters of Fig.~(a). 
(c) Momentum distribution $n_\sigma(k)$ of Eq.~\eqref{eq:H_eff_phi} for $\rho=0.5$, and $\phi=\pi/2$ ($L=60$).}
\label{fig:pd_dJFH}
\end{figure*}


Although for $J(t)\ll \Delta, |\Delta\pm U|$ direct hoppings are energetically forbidden, virtual hoppings may induce effective interactions between NN sites~\cite{Suppl} of the form
\begin{eqnarray}
\HHop_{NN}&=& \!\! \sum_{\langle i,j \rangle} \left [  \frac{2J_0^2}{\Delta+U} P^0_i P^2_j -\frac{2J_0^2}{\Delta-U} P^2_i P^0_j \right \delimiter 0 \nonumber\\
&+&\!\!\!\frac{J_0^2}{\Delta} \left( (1-n_i) P^1_j - P^1_i (1-n_j) \right)\\
&+&\!\!\!\left\delimiter 0 \frac{2 UJ_0^2}{\Delta^2 - U^2} ( P^{1\uparrow}_i P^{1\downarrow}_j \! + \! P^{1\downarrow}_i P^{1\uparrow}_j \! - \!S^+_i S^-_j \! - \! S^-_i S^+_j)\right ],  \nonumber
\label{eq:H_2nd}
\end{eqnarray}
where $S^+_i\!=\!c_{i,\uparrow}^\dagger c_{i,\downarrow}$ and $S^-_i\!=\!c_{i,\downarrow}^\dagger c_{i,\uparrow}$ are spin operators, $n_i\! =\! n_{i,\downarrow} \!+\! n_{i,\uparrow}$, and we introduce the projector of two particles per site $P^2_i\!=\!n_{i,\downarrow} n_{i,\uparrow}$, zero particles $P^0_i\!=\!(1- n_{i,\downarrow})(1-n_{i,\uparrow})$, and a single particle $P^{1\sigma}_i \!=\! (1- n_{i,\bar\sigma})n_{i,\sigma}$, and $P^1_i=P^{1\downarrow}_i+P^{1\uparrow}_i$. The peculiar NN interactions depend on $J_0^2/\Delta$ and $J_0^2/(U\pm \Delta)$, whereas the 
effective hopping is given by $\delta J_{i}$. Hence they may be separately controlled. For sufficiently small $J_0\!\ll\! \Delta, |U\pm \Delta|$ we 
may neglect $\HHop_{NN}$. However, as shown below, $\HHop_{NN}$ opens additional interesting possibilities.



\paragraph{Non-equilibrium dynamics.--} Figure~\ref{fig:2}(a) depicts our results for 
the dynamics of the averaged probability of double occupancy, $\langle P_2 \rangle$, based on exact diagonalization of small systems~\cite{footnote-numerics}. 
We initially prepare for $\delta V=0$ a Mott-insulator~(MI) state at $U\gg J_0$, 
assuming an initial temperature  $T = J_0$, and hence initially $\langle P_2\rangle \simeq 0$.  At time $t=0$ we abruptly turn on the modulation $\delta J(t)$. 
The results show a very good agreement between the effective model $\HHop_{eff}+\HHop_{NN}$, and the full model~\eqref{eq:H_full_J(t)}. Figure~\ref{fig:2}(a) shows that 
non-equilibrium experiments should be able to reveal both the ODG, and the suppression of $\langle P_2 \rangle$ resulting from the repulsive effective interactions $\tilde{U}$.

The analysis of ground-state properties requires a (quasi-)adiabatic ramping of $\delta V$. 
We present in Fig.~\ref{fig:2}(b) our results obtained using infinite time evolving block decimation (iTEBD)~\cite{Vidal07}. 
Starting again with $\delta V=0$ from an initial MI state, we have studied 
the quasi-adiabatic preparation of different MI states. During the time $0<t<t_{\rm ramp}$ we linearly increase $\delta V$ to its final value, monitoring $\langle P_2 \rangle$. 
Again $\HHop_{eff}+\HHop_{NN}$ reproduces very well the dynamics of the full model~\eqref{eq:H_full_J(t)}.
After the ramp, the heating induced by the quasi-adiabatic character of the finite ramping time results in oscillations of $\langle P_2 \rangle$ around the 
value expected for the ground state of the effective model~(see below).


\paragraph{Phases of the effective Hamiltonian.--} At this point we focus on the ground-state physics of $\HHop_{eff}$, assuming that $J_0\ll \Delta, |\Delta\pm U|$, and hence that $\HHop_{NN}$ may be neglected.
For $\beta\neq 1$, $\HHop_{eff}$ realizes a broad class of Hubbard models with correlated hopping extensively studied in the context of cuprate 
superconductors~\cite{Arrachea1994,Arrachea1996,Arrachea1997,Aligia1999,Aligia2000}, and recently revisited for ultra-cold gases with modulated interactions~\cite{DiLiberto2014,Greschner14doubleshaking}. 
For $\phi\neq 0$, the ODG gives rise to a particularly intriguing physics. 
For $\beta=1$: 
\begin{equation}
\HHop_{eff}=-\frac{\delta J_1}{2}\sum_{\sigma,j} c_{j+1,\sigma}^\dag {\rm e}^{\ii \phi |n_{\bar\sigma, j+1}-n_{\bar\sigma,j}|} c_{j,\sigma}+\tilde U \HHop_{int}.
\label{eq:H_eff_phi}
\end{equation}
For a low lattice filling $\rho$ for which processes (iv) may be neglected,  
a Jordan-Wigner like transformation~\cite{Keilmann2011}, $f_j=\e^{\ii 2 \phi \sum_{1\leq l< j} n_l} \e^{\ii \phi n_j} c_j$, maps~\eqref{eq:H_eff_phi} into a two-component anyon-Hubbard model~(2-AHM):
\begin{eqnarray}
\!\!\!\HHop_{AHM}\! \! &=&\! \!  - \frac{\delta J_1}{2} \sum_{j,\sigma} (f_{j,\sigma}^{\dagger}f_{j+1,\sigma}^{\phantom \dagger} \!+\! \text{H.c.}) \!+\! \tilde{U}\HHop_{int} \,.
\label{eq:H_anyons}
\end{eqnarray}
where the operators $f_{j,\sigma}$ and $f_{j,\sigma}^\dagger$ characterize anyon-like hardcore particles that fulfill a deformed exchange statistics~(DES): 
$f_{j,\sigma} f_{k,\sigma'} + \mathcal{Q}^{\sigma,\sigma'}_{j,k} f_{k,\sigma'}^\dagger f_{j,\sigma} = \delta_{jk} \delta_{\sigma,\sigma'}$ 
and $f_{j,\sigma} f_{k,\sigma'} + \mathcal{Q}^{\sigma,\sigma'}_{j,k} f_{k,\sigma'} f_{j,\sigma} = 0$, 
with $\mathcal{Q}^{\sigma,\sigma'}_{j,k} = \e^{\ii 2 \phi}$~($j>k$), 
$0$~($j=k$), $\e^{-\ii 2 \phi}$~($j<k$). Specific cases of the 2-AHM have been studied in the context of exactly solvable models~\cite{Schulz1998,Osterloh2000}. 
In contrast, the non-integrable DES discussed here does strongly modify the spectrum of the 2-AHM compared to the fermionic Hubbard model. 

Figure~\ref{fig:pd_dJFH}(a) shows, as a function of $\beta$ and the chemical potential $\mu$, the ground-state phase diagram of~\eqref{eq:H_eff_phi} for $\phi=\pi/2$ and $\tilde U=0$, obtained by means of  
density matrix renormalization group (DMRG)~\cite{Schollwock2011} simulations in finite-size open-boundary systems of up to $80$ sites, keeping up to $600$ matrix states~\cite{footnote-density}.
For $\beta=0$ doubly-occupied sites~(doublons) and empty ones~(holons) become mutually impenetrable, resulting at half filling 
in a non-conducting metal with a vanishing Drude weight~(Kohn metal)~\cite{Arrachea1996}.
For $0<\beta<1$,  in the absence of ODG, the system undergoes a smooth phase transition from a metal~(M) with dominant spin-density wave~(SDW) correlations, $(-1)^j\langle n_{0-}n_{j-}\rangle$, 
with $n_{j-}=n_{j,\uparrow}-n_{j,\downarrow}$, to a triplet superconductor~\cite{Greschner14doubleshaking}. On the contrary, for $\phi=\pi/2$, the M phase undergoes 
for $\beta\lesssim 1.4$ a commensurate-incommensurate~(C-IC) phase transition,  marked by a kink in the $\mu(\rho)$ curve~(Fig.~\ref{fig:pd_dJFH}(b)), to a peculiar gapless multi-component~(MC) phase. 
The MC phase presents a central charge $c\approx 3$~\cite{footnote-CFT,Vidal2003, Calabrese2004}. In contrast, the metallic phase has $c=2$. 
The MC phase smoothly connects to the Kohn-metal for $\beta\to 0$.
For $\beta \gtrsim 1.4$ and $\rho\neq 1$, a spin gap $\Delta_S$ opens and the kink in $\mu(\rho)$ disappears marking the transition to a 
phase with dominant singlet-superconducting~(SS) correlations, $\langle {\cal Q}_{0-}^\dag {\cal Q}_{j-}\rangle$, 
with ${\cal Q}_{j-}\equiv c_{j+1,\downarrow}  c_{j,\uparrow} - c_{j+1,\uparrow}  c_{j,\downarrow}$.
Finally, at $\rho=1$  we find a MI with dominant SDW correlations, and a totally gapped phase with bond-ordering wave~(BOW) order
$\mathcal{O}_D(x) = \sum_\alpha T_\alpha(x)-T_\alpha(x+1)$, with $T_\alpha(x) = c_{\alpha,x}^\dagger c_{\alpha,x+1} + H.c.$.


\paragraph{MC phase.--} The MC phase, which occurs even for $\beta=1$ and $\tilde U=0$, is a direct consequence of the ODG. 
The nature of this phase is best understood for $\phi=\pi/2$ and $\beta=1$. 
In that case, the two-particle problem, with a $\uparrow$ particle and a $\downarrow$ one, presents for any $\tilde U$ 
an exact bound eigenstate, $|P\rangle=\cos\theta |D\rangle+\ii\sin\theta |S\rangle$, with energy $E_P=\frac{\tilde U}{2}\!-\!\sqrt{\frac{\tilde U^2}{4}\!+\! 2{\delta J_1}^2}$, where 
$\tan\theta\!=\!\frac{\tilde U\!-\! E_P}{\sqrt{2} \delta J_1}$, $|D\rangle\!=\!\sum_j (-1)^j |\!\!\uparrow,\downarrow\rangle_j$, and 
$|S\rangle\! =\!\sum_j (-1)^j (|\!\!\uparrow\rangle_j |\!\!\downarrow\rangle_{j+1}\!\!-|\!\!\downarrow\rangle_j |\!\!\uparrow\rangle_{j+1})/\sqrt{2}$.
The existence of this bound state even for $\tilde U>0$ results from the ODG~(see Suppl. Material~\cite{Suppl}).
For sufficiently large $\tilde U>0$, $E_P>2E_F$, with $E_F$ the Fermi energy of the metal, and the M phase is stable. 
For decreasing $\tilde U$, $E_P<2E_F$, and part of the Fermi sea forms pairs that quasi-condense in $|P\rangle$, until the 
new Fermi energy $E'_F=E_P/2$.  The MC phase results from the coexistence of a partially depleted Fermi sea and bound pairs. 
When $E'_F$ reaches the bottom of the lattice band, the Fermi sea is fully depleted marking the onset of the SS phase.

The MC phase has a characteristic momentum distribution of both components, $n_\sigma(k)$, and 
it can be thus easily revealed in time-of-flight measurements. Figure~\ref{fig:pd_dJFH}(c) shows our results for $n_\sigma(k)$ for $\phi\!=\!\pi/2$. For large-enough $\tilde U$ the M phase presents a 
slab-like Fermi sea. In the MC phase, the slab shrinks due to partial pairing. The latter results in a blurred contribution to $n_\sigma(k)$, 
$\frac{1}{2\pi}\! \left [ 1\!-\!\sqrt{2}\sin(2\theta)\sin(k/2)\!-\!\sin^2\theta \cos(2k) \right ]$, as expected for $|P\rangle$ pairs~\cite{Suppl}. 
The MC-SS transition is marked by the vanishing Fermi sea.



\begin{figure}[t]
\begin{center}
\includegraphics[height=4.5cm]{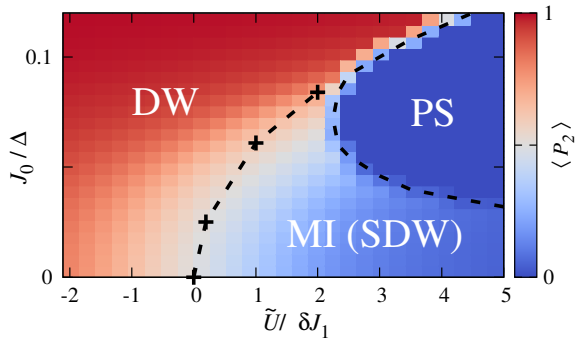}
\caption{(Color online) Phase diagram of $\HHop_{eff}+\HHop_{NN}$ for $\rho=1$, $U=5 J_0$, $\delta J=0.1 J_0$, $\beta=1$, and $\phi=0$. 
The MI-DW transition is given by $K_S=1$~(extrapolated from DMRG calculations of up to $L=160$ sites~\cite{footnote-KS}). 
The coloring codes $\langle P_2 \rangle$ obtained from iTEBD calculations~(for $200$ states results are consistent with our DMRG data of $160$ sites).}
\vspace*{-0.5cm}
\label{fig:pd2nd}
\end{center}
\end{figure}



\paragraph{Intersite interactions.--} $\HHop_{NN}$ becomes relevant for large-enough $J_0/\Delta$, $J_0/|\Delta\pm U|$. Combining effective on-site and NN interactions 
constitutes an additional interesting control possibility resulting from the 3CM. Figure~\ref{fig:pd2nd} depicts for $\beta\!=\!1$, $\phi\!=\!0$ 
and $\rho\!=\!1$
the phase diagram as a function of $\tilde U/\delta J$, and $J_0/\Delta$~(which controls the strength of the NN terms). For $J_0/\Delta\to 0$ the 
standard two-component 1D Fermi Hubbard model is recovered~\cite{EsslerBook}. For any $\tilde{U}\!>\!0$ there is a MI~(SDW) phase 
with a finite charge excitation gap $\Delta_c\!>\!0$ but $\Delta_S\!=\!0$, whereas for $\tilde{U}\!<\!0$ $\Delta_S\!>\!0$ and $\Delta_c\!=\!0$. 
For sufficiently large $J_0/\Delta$ the system is driven into a fully gapped density-wave~(DW), characterized by a non-vanishing 
DW order $(-1)^j \langle n_{0} n_{j} \rangle$. 
For $\tilde U>0$ we observe two MI phases with a suppressed doublon number, the above mentioned MI~(SDW) 
and a region of phase separation of ferromagnetic domains~(PS). 
The MI-DW transition is associated to the opening of $\Delta_S$, characterized by the Luttinger-liquid parameter in the spin sector $K_S=1$~($+$-symbols in Fig.~\ref{fig:pd2nd})~\cite{footnote-KS,Moreno2011}.  
Since $\HHop_{NN}$ breaks the spatial reflection symmetry, 
we do not observe a separate BOW phase, as it is the case for Hubbard models with standard density-density NN interactions~\cite{Ejima07}, but a non-zero 
BOW-order in the DW due to the preferred creation of excitations in a particular spatial direction.



\paragraph{Outlook.--} A multicolor modulation of the lattice depth allows for a flexible separate manipulation of (a) correlated hopping, 
controlled by the modulation amplitudes $\delta V_s$; (b) ODG, given by the dephasings $\phi_s$; 
(c) effective on-site interactions, provided by the detuning $\tilde U$; and (d) NN interactions, that depend on $J_0/\Delta$ and $J_0/|\Delta\pm U|$. 
3CM thus provides an experimentally straightforward method for engineering a very broad class of lattice models, 
including Hubbard Hamiltonians with correlated hopping, peculiar extended models, and 2-AHM. In particular, the controllable 
quantum statistics of the 2-AHM results in a peculiar MC phase of coexisting superconducting and metallic components. 
The RWA requirements necessary for the 3CM are readily achievable experimentally. For example,
for $^{173}{\rm Yb}$~(scattering length of $199.4 a_B$ and lattice spacing of $380{\rm nm}$~\cite{Mancini2015,Hofer2015}) with $s_0\!=\! 6.9$~($J_0/h\!=\!100$Hz), 
one achieves  $U\!=\!23J_0$, $\Delta\!=\!16J_0$, $|\Delta\!-\!U|\!=\! 7J_0$, well within the RWA requirements.
For $\delta J/J_0=0.2$,  the typical effective-tunneling time is $\tau=\hbar/\delta J \simeq 8$ ms.

Multi-color modulation permits several further interesting extensions, including the control of three-body interactions~\cite{Daley14}. In combination with a Raman-induced coupling of several spin components~\cite{Mancini2015, Stuhl2015}
one may study density dependent magnetic fields~\cite{Greschner15DDSM}. 
Other scenarios could pave a realistic exploration path towards the simulation of dynamical gauge fields with cold atoms in optical lattices, exploring e.g.  
occupation-dependent non-Abelian fields and gauge fields in Fermi-Bose mixtures.


\begin{acknowledgments}
We thank L. Fallani, A. Eckardt, and C. de Morais-Smith for discussions. 
We acknowledge support of QUEST-LFS and the DFG Research Training Group 1729.
Simulations were carried out on the cluster system at the Leibniz University of Hannover, Germany.
\end{acknowledgments}


\bibliographystyle{prsty}

\begin{thebibliography}{}
\bibitem{footnote-numerics} Similar agreement between the dynamics of the full model and the effective one is obtained in the thermodynamic limit 
employing infinite time evolving block decimation (iTEBD) simulations~\cite{Vidal07},  
which are possible using the translational invariant formulation of Eq.~\eqref{eq:H_trans_inv}.
\bibitem{footnote-density} Hamiltonian~\eqref{eq:H_eff} is symmetric under particle-hole exchange, and hence the region $\mu\! >\! 0$ ($\rho \! >\! 1$) is mirror symmetric to that 
depicted for $\mu\! <\! 0$ ($\rho\! < \! 1$) in Fig.~\ref{fig:pd_dJFH}(a).
\end{thebibliography}

\end{document}


\title{Supplementary material for "Engineering interactions and anyon statistics by multicolor lattice-depth modulations"}

\author{Lorenzo Cardarelli} 
\affiliation{\hannover}
\author{Sebastian Greschner} 
\affiliation{\hannover}
\author{Luis Santos}
\affiliation{\hannover}

\begin{abstract}
In this Supplementary Material we discuss in more detail about the physics of the multi-component~(MC) phase. 
We comment as well in more detail about the derivation of the effective Hamiltonian using a Magnus expansion.
\end{abstract}

\maketitle

\section{Multi-component phase}

\subsection{Two-particle model}

We assume for simplicity $\beta=1$, and hence Model (5) of the main text.  
We are interested in the two-particle problem, with one $\uparrow$ particle and one $\downarrow$ particle.
Let $|D(j)\rangle$ be a doubly ocuppied site ($j$ site), and $|S(j,j+l)\rangle$ a singlet state placed in sites $j$ and $j+l$.
Then 
\begin{eqnarray}
\HHop_{eff} |D(j)\rangle&=&-\frac{\delta J_1}{\sqrt{2}} \left [ e^{\ii\phi} |S(j,j+1)\rangle +e^{-\ii\phi} |S(j-1,j)\rangle \right ] + \tilde U |D(j)\rangle,  \\
\HHop_{eff} |S(j,j+1)\rangle&=&-\frac{\delta J_1}{\sqrt{2}} \left [ e^{\ii\phi} |D(j+1)\rangle +e^{-\ii\phi} |D(j)\rangle \right ]- \frac{\delta J_1}{2} \left [ |S(j-1,j+1)\rangle + |S(j,j+2)\rangle \right ], \\
\HHop_{eff} |S(j,j+l)\rangle&\myeq&-\frac{\delta J_1}{2} \left [ |S(j-1,j+l)\rangle + |S(j+1,j+l)\rangle + |S(j,j+l-1)\rangle + |S(j,j+l+1)\rangle \right ].
\end{eqnarray}
Let $|D(k)\rangle=\frac{1}{\sqrt{L}}\sum_l e^{\ii kl}|D(l)\rangle$ and $|S(j,k)\rangle=\frac{1}{\sqrt{L}}\sum_l e^{\ii k(l+j/2)}|S(l,l+j)\rangle$, with 
$k$ the center-of-mass momentum of the pair, and $L$ the number of sites. Then $\HHop_{eff}=\sum_k \HHop_{eff}(k)$, with $\HHop_{eff}(k) =\HHop_0(k)+\HHop_1(k)$, where:
\begin{eqnarray}
\HHop_0(k)&=&\tilde U |D(k)\rangle\langle D(k)| - A(k) \left [|S(1,k)\rangle\langle D(k)| + h.c. \right ],  \\
\HHop_1(k)&=&- B(k) \sum_{j\ge 1} \left [|S(j,k)\rangle\langle S(j+1,k)| + h.c. \right ],
\end{eqnarray}
with $A(k)=\sqrt{2} \delta J_1 \cos(k/2-\phi)$ and $B(k)=\delta J_1 \cos(k/2)$.
We may diagonalize $\HHop_0$:
\begin{equation}
\HHop_0(k)=E_+(k) |\tilde P(k)\rangle\langle \tilde P(k)| + E_-(k) |P(k)\rangle\langle P(k)|, \\
\end{equation}
where the eigenenergies are $E_\pm (k)=\frac{\tilde U}{2}\pm \sqrt{\left ( \frac{\tilde U}{2}\right )^2+A(k)^2}$,
and the corresponding eigenstates are $|\tilde P(k)\rangle=\cos\theta(k) |D(k)\rangle+\sin\theta(k)|S(1,k)\rangle$, and $|P(k)\rangle=-\sin\theta(k) |D(k)\rangle+\cos\theta(k)|S(1,k)\rangle$, with 
$\tan \theta(k)=\frac{\tilde U/2-E_-(k)}{A(k)}$. The Hamiltonian $\HHop_0$ characterizes deeply-bound pairs.
We may then split $\HHop_1(k)=\HHop_c(k)+\HHop_u(k)$, where   
\begin{equation}
\HHop_u(k)= -B(k) \sum_{j\ge 2} \left [|S(j,k)\rangle\langle S(j+1,k)| + h.c. \right ]
\end{equation}
determines the physics of broken pairs, where the dynamics of relative coordinate $j$ is given by the hopping rate 
$B(k)$, and 
\begin{equation}
\HHop_c(k)=-B(k) \left ( \sin\theta(k) |P(k)\rangle +\cos\theta(k) |\tilde P(k)\rangle \right ) \langle S(2,k)| + h.c,   \\
\end{equation}
characterizes the coupling between deeply-bound and unbound pairs. Note that such a coupling is also given by $B(k)$.
 
Let us consider $\phi=\frac{\pi}{2}$. 
In that case, $E_\pm(k)=\frac{\tilde U}{2}\pm \sqrt{\left ( \frac{\tilde U}{2}\right )^2+2\delta J_1^2\sin^2 (k/2)}$. The minimal energy 
is clearly for $k=\pi$, $E_P\equiv E_-(\pi)=\frac{\tilde U}{2}\pm \sqrt{\left ( \frac{\tilde U}{2}\right )^2+2\delta J_1^2}$. If existing, bound pairs will quasi-condense in $|P\rangle\equiv |P(\pi)\rangle$. 
Crucially, $B(\pi)=0$, and hence $\HHop_c=0$. As a result, $|P\rangle$ remains a deeply-bound two-particle eigenstate, fully decoupled from the unbound pairs, irrespective of the value of $\tilde U /\delta J_1$. 
On the contrary for  $\phi=0$, i.e. without occupation-dependent gauge (ODG), the bound pairs 
are fully connected with the rest and cannot be formed unless $\tilde U<0$ dominates. For $\phi$ in the vicinity of $\pi/2$ the coupling $\HHop_c$ may be considered perturbative, and 
deeply-bound pairs due to the ODG still exist even if $\phi$ is not exactly $\pi/2$.

The existence of these pairs  that are deeply-bound by the ODG rather than by attractive interactions is crucial to understand the nature of the MC phase.
The metallic~(M) phase is stable if $E_P/2>E_F$, with $E_F$ the Fermi energy of the metal. However, for decreasing $\tilde U>0$, $E_F<E_P/2$, and hence 
it is energetically favorable to pair part of the Fermi sea into $|P\rangle$ pairs, until reaching an equilibrium at a new Fermi energy $E'_F=E_P/2$. 
This partial pairing, and the corresponding coexistence of a two-component metal and a superconductor explains the MC phase, and its $c=3$ central charge.
For  $E_-(\pi)<-2\delta J_1$~(which occurs at $\tilde U/\delta J_1\simeq -1$)  the Fermi sea is completely depleted, and the system enters the fully-paired~(SS) phase.

\subsection{Momentum distribution}

The momentum distribution of the $\uparrow$ component in the $|P\rangle$ state is $n_\uparrow^{(P)}(k)=\sum_{i,j}e^{\ii k(i-j)}\langle P| c_{i,\uparrow}^\dag c_{j,\uparrow} |P\rangle$, where 
\begin{eqnarray}
\langle P|c_{l,\uparrow}^\dag  c_{l,\uparrow} |P\rangle &=&   \frac{1}{L}, \\
\langle P|c_{l+1,\uparrow}^\dag  c_{l,\uparrow} |P\rangle=\langle P|c_{l-1,\uparrow}^\dag  c_{l,\uparrow} |P\rangle^* &=&   \frac{-\sin(2\theta(\pi))}{L\sqrt{2}}e^{\ii\pi/2}, \\
\langle P|c_{l+2,\uparrow}^\dag  c_{l,\uparrow} |P\rangle=\langle P|c_{l-2,\uparrow}^\dag  c_{l,\uparrow} |P\rangle &=&  \frac{-\sin^2(\theta(\pi))}{2L}, \\
\end{eqnarray}
and other correlations are zero. After normalizing:
\begin{equation}
n_\uparrow^{(P)}(k)=\frac{1}{2\pi} \left [ 1-\sqrt{2}\sin(2\theta(\pi))\sin(k/2)-\sin^2\theta(\pi)\cos(2k) \right ]
\end{equation}
with $\theta(\pi)=\arctan \left  [\chi + \sqrt{\chi^2+1} \right ]$, with $\chi=\frac{\tilde U}{2\sqrt{2} \delta J_1}$. For the $\downarrow$ component the expression is identical.
This expression is in excellent agreement with the 
blurred momentum distribution that is found in our numerics in the MC phase~(Fig. 3(c) of the main text) in addition to the partially-depleted slab-like Fermi sea.

\section{Derivation of the effective model via Magnus expansion}

For the simplified case of a time periodic Hamiltonian, i.e. assuming that the frequencies $\Delta+U$ and $\Delta$ are integer multiples of $\omega\equiv \Delta-U$, we may obtain the same effective Hamiltonian of Eqs.~(3) and (4) of the main text employing a formal Magnus expansion~\cite{maricq1982, jotzu2014, goldman2014periodically, aidelsburger2015chern} or Floquet analysis~\cite{Straeter16}. 
Following the presentation of Ref.~\cite{aidelsburger2015chern} we may express the effective Hamiltonian as a series in $1/\omega$ as
$\HHop_{eff} = \HHop^{(0)} + \HHop^{(1)}_{ME} + \mathcal{O}\left(\frac{1}{\omega^2}\right)$.
The lowest order term
\begin{align}
\HHop^{(0)} = \frac{1}{T} \int_0^{T} dt_1 \HHop(t_1)
\end{align}
provides Eq.~(3) of the main text. The first order correction in $\frac{1}{\omega}$ may be expressed as~\cite{maricq1982} 
\begin{align}
\HHop^{(1)}_{ME} = \frac{-\ii}{2 T} \int_0^{T} dt_2 \int_{0}^{t_2} dt_1 [ \HHop(t_2), \HHop(t_1)].
\end{align}
If the time periodic Hamiltonian is given by a Fourier series $\HHop(t) = \HHop_0 + \sum V^{(k)} \e^{i k \omega t}$, then
\begin{align}
\HHop^{(1)}_{ME} &= \frac{1}{\omega} \sum_k \frac{1}{k} \left( [V^{(k)}, V^{(-k)}] - [V^{(k)}, \HHop_0] + [V^{(-k)}, \HHop_0] \right).
\label{eq:01_lattice_depth_eff_H1}
\end{align}
In Eq.~(2) of the main text we expand the exponential term $\e^{\pm\ii t Un_{j\sigma}} = 1 + (\e^{\pm\ii t U}-1) n_{j\sigma}$. Then
\begin{align}
\tilde{\mathcal{H}}(t) = (J_0+\delta J(t))\left( \e^{{\rm i}t \left[\Delta - U\right]} \bar{V}^{(1)} + \e^{{\rm i}t \Delta} \bar{V}^{(2)}
+\e^{{\rm i}t \left[ \Delta +U\right]} \bar{V}^{(3)} + {\rm H.c.} \right)
\label{eq:H_trans_inv_exp}
\end{align}
with
\begin{align}
\bar{V}^{(1)} &= \sum_{j,\sigma} d_{j,\sigma}^\dagger c_{j+1,\sigma} - d_{j,\sigma}^\dagger d_{j+1,\sigma}\,,\nonumber\\
\bar{V}^{(2)} &= \sum_{j,\sigma} \left(d_{j,\sigma}^\dagger - c_{j,\sigma} \right) \left(d_{j+1,\sigma}^\dagger - c_{j+1,\sigma} \right)\,,\nonumber\\
\bar{V}^{(3)} &= \sum_{j,\sigma} c_{j,\sigma}^\dagger d_{j+1,\sigma} - d_{j,\sigma}^\dagger d_{j+1,\sigma} \,.
\label{eq:H_V}
\end{align}
where we employ the correlated annihilation operator $d_{j,\sigma} \equiv n_{j,\bar{\sigma}} c_{j,\sigma}$.
Neglecting terms of order $J_0 \delta J$ and $\delta J^2$ we may write
\begin{align}
\HHop^{(1)}_{ME} = \frac{J_0^2}{\Delta-U} \left[\bar{V}^{(1)}, \bar{V}^{(1)\dagger}\right] + \frac{J_0^2}{\Delta} \left[\bar{V}^{(2)}, \bar{V}^{(2)\dagger}\right] 
+\frac{J_0^2}{\Delta+U} \left[\bar{V}^{(3)}, \bar{V}^{(3)\dagger}\right] + \mathcal{O}\left(\delta J\right), 
\end{align}
which after some algebra yields Eq.~(4) of the main text.

\bibliographystyle{prsty}